\documentclass[accepted]{smartcom_ww}
\usepackage{latexsym}

\usepackage{cite}
\usepackage[dvips]{graphics}
     \usepackage{leqno}
     \usepackage{color}
     \usepackage{graphicx}
    \usepackage[cmex10]{amsmath}
     \usepackage{amssymb}
     \usepackage{graphicx}
     \usepackage{multicol}
     \usepackage{flushend}

\etitle{\huge{Cooperative Beamforming in Cognitive Radio Relay Networks Using Amplify-and-Forward Relaying Technique}}
\authorlist{%
 \authorentry[amir.behrouzifar@rutgers.edu]{}{Amir Behrouzi-Far}{Rutgers}%
 \authorentry[s\_mohammadkhani@elec.iust.ac.ir]{}{Saeideh Mohammadkhani}{IUST}%
}
\affiliate[Rutgers]{}{Department of Electrical and Computer Engineering , Rutgers University, Piscataway, NJ 08854, USA}
\affiliate[IUST]{}{Departmetn of Electrical Engineering, Iran Unicersity of Science and Technology, Tehran, Iran}

\begin{document}
\begin{eabstract}
In this work, we study the problem of relay beamforming for an underlay cognitive radio relay network using amplify-and-forward (AF) relaying technique. We consider a cognitive radio network consisting of single primary and multiple secondary transmitters/receivers. In addition, there are several relay nodes that help both primary and secondary networks. We propose a new beamforming method for relays' transmission, in which the beamforming weights at the relays are the solution of an optimization problem. The objective of the optimization problem is maximizing the worst case signal to interference plus noise ratio (SINR) at the secondary receivers (SU-RX) and the constraints are the interference power on the primary receiver (PU-RX) and the total transmitted power of the relay nodes. We show that the beamforming problem can be reformulated as a second order cone programming (SOCP) problem and solved by the bisection search method. Our simulation results show the performance improvement as a function of the size and the power of the relay network.
\end{eabstract}
\begin{ekeyword}
Cognitive Radio, Spectrum Sharing, Relay, Beamforming, SOCP.
\end{ekeyword}
\maketitle

\section{Introduction}
By introducing the secondary users (SU), which coexist in the same frequency band with some primary users (PU), cognitive radio (CR) technology can improve spectral efficiency \cite{goldsmith2009breaking}. Depending on how the frequency band of the PUs are utilized by the SUs, three operating modes for CR networks (CRN) exists; overlay, underlay and interweave \cite{goldsmith2009breaking}.

The promise of improving spectral efficiency comes with several challenges. Among all, providing enough signal to interference plus noise ratio (SINR) to SU-RX and limiting the SU interference on the PU-RX have attracted major attention \cite{zheng2010robust,piltan2012distributed,du2011joint}. Beamforming techniques, implemented on the transmitter antennas, are among the most studied solutions for the aforementioned challenges \cite{chen2009filter,mohammadkhani2014robust,mohammadkhani2012cooperative}.

Due to the size and power limitations on SU-TX, beamforming can be done by a relay network \cite{mohammadkhani2012cooperative}. Relay assisted cognitive radio networks have been studied in the literature. Authors in \cite{beigi2009cooperative}, considered  a single-primary single-secondary CR, and studied the beamforming problem with the objective of maximizing SU-RX SINR. They employed a genetic algorithm to solve the optimization problem. Using SUs as the relay nodes for PUs is proposed in \cite{bayat2011cognitive}. They proposed a distributed spectrum access algorithm, constrained with the minimum sum rate of the network.

In this paper, we consider a scenario where $M$ secondary source-destination pairs use the same frequency band of a primary transmitter-receiver pair, in an underlay mode. In addition, there are some relay nodes that help both primary and secondary networks. We propose a cooperative beamforming method which is performed by the relay network. The objective of the beamforming is to maximize the worst case SINR in the secondary receivers, while the interference on the PUs stays less than a limit and the total power of the relay network is fixed. Designing the beamforming weights leads to an optimization problem, which we reformulate it as a SOCP problem. Since the problem is convex, we employed bi-section search method to find the optimal value \cite{boyd2004convex}. Our simulation results show that for a given transmitted power of the relay network, the system performance is improved as the number of relays increases. Moreover, the effect of the number of the SUs and interference threshold on the PU-RX is also studied in our simulations.

The rest of this paper is organized as follows. In Section 2 we introduce the system model and formulate the problem. We present the beamforming design through optimization problem in Section 3. In Section 4 we present simulation results. Concluding remarks are presented in Section 5. 

\emph{Notations:} superscripts $(.)^T$, $(.)^*$ and $(.)^H$ denote transpose, conjugate and conjugate transpose, respectively. $diag(A)$ is a diagonal matrix with the vector $A$ being its diagonal entries. $E(x)$ represents the expectation value of the random variable $x$ and $||.||_2$ denotes the second norm.\\

\section{System Model and Problem Formulation}
\begin{figure}[t]
   \centering
    \includegraphics[width=\columnwidth]{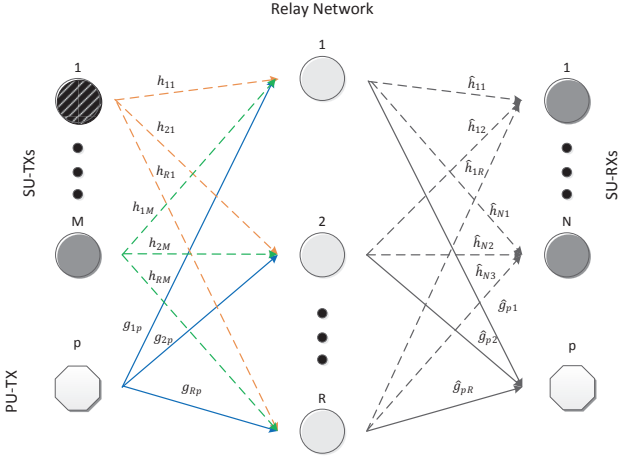}
    \caption{System model.}\label{figsystemmodel}
   \end{figure}
As shown in Fig. 1, we consider a cognitive radio relay network in which $M$ secondary transmitters communicate with $N$ secondary receivers in the same frequency band with a pair of primary transmitter-receiver. A relay networks, with $R$ nodes also co-exist with the CRN. A communication between the primary network and the secondary network takes place in two phases. In the first phase, the transmitters in CRN send communicate their signals to the relay nodes. The recieved signal at the relay nodes can be written as,
\begin{equation}\label{eq01}
{y^{(1)}_r} = \sqrt {{P_p}} {g_{rp}}x_p + \sqrt {{P_s}} {{\bar h}_r^T}\bar x_s^{(1)} + {n_r}, \hspace{5mm} r=1,2,...,R
\end{equation}
in which,
\begin{equation}{}\nonumber
{{\bar h}_r} = [{h_{r1}}{\rm{ }},{h_{r2}}{\rm{ }},\ldots,{\rm{ }}{h_{rM}}]^T\nonumber,\\
{\bar x_s^{(1)}} = {\left[ {{x_1^{(1)}}{\rm{ }},{x_2^{(1)}}{\rm{ }},\ldots,{\rm{ }}{x_M^{(1)}}} \right]^T}\nonumber,
\end{equation}
where $x_i$ and $h_{ri}$, $i=1,2,...,M$, indicate the transmitted symbol from the $i$th SU-TX and the channel gain between the $i$th SU-TX and the $r$th relay, respectively. ${x_{p}}$ and $g_{rp}$ indicate the PU-TX transmitted symbol and channel coefficient between the PU-TX and the $r$th relay. In this model, the transmitted symbols are power normalized. Finally, $n_{r}$ is zero mean Additive White Gaussian Noise (AWGN) with variance ${\sigma^2_n}$. We assume that there is no direct link between transmitters and receivers and, therefore, the receivers do not receive any signal during the first time slot.

Using AF technique , the relay nodes multiply the received signals by some beamforming weights and communicate them to the receivers in the second phase. Accordingly, the transmitted signal from the $r$th relay could be written as,
\begin{equation}
{x_r^{(2)}} = {\omega_r}\frac{{{y^{(1)}_r}}}{{\sqrt {{P_p}{{\left| {{g_{rp}}} \right|}^2} + {P_s}\bar h_r^H{{\bar h}_r} + \sigma _n^2} }}, \hspace{5mm} r=1,2,...,R
\end{equation}
in which $\omega_{r}$ is the beamforming weight in the $r$th relay. The received signal at the $j$th SU-RX, in the second phase, is
\begin{equation}
{y^{(2)}_j} ={ \overline {\hat h}_j}^T {\overline x_r^{(2)}} + {n_j}, \hspace{5mm} j=1,2,...,N
\end{equation}
in which,
\begin{equation}{}\nonumber
{ \overline {\hat h}_j}=[\hat h_{j1},\hat h_{j2},...,\hat h_{jR}]^T\nonumber
{\bar x_r^{(2)}} = {\left[ {{x_1^{(2)}}{\rm{ }},{x_2^{(2)}}{\rm{ }},\ldots,{\rm{ }}{x_M^{(2)}}} \right]^T}\nonumber,
\end{equation}
where $\hat h_{jr}$ indicates the channel coefficient between the $r$th relay and the $j$th SU-RX and $n_{j}$ is zero mean AWGN with variance ${\sigma^2_j}$. Substituting $(1)$ and $(2)$ in $(3)$, we have
\begin{equation}{}\nonumber
{y^{(2)}_j} = \mathop \sum \limits_{r=1}^R {\omega _r}{\hat h_{jr}}\frac{{\sqrt {{P_p}} {g_{rp}}x_p + \sqrt {{P_s}} {\bar h_r^T}{\bar x_s} + {n_r}}}{{\sqrt {{P_p}{{\left| {{g_{rp}}} \right|}^2} + {P_s}\bar h_r^H{{\bar h}_r} + \sigma _n^2} }}\\ \hspace{10mm} + {n_j},
\end{equation}
which could be decomposed into three parts
\begin{subequations}\
\begin{equation}{}\nonumber
\begin{split}
{y^{(2)}_{j_{(p)}}} &= \mathop \sum \limits_{r=1}^R {\omega _r}{\hat h_{jr}}\frac{{\sqrt {{P_p}} {g_{rp}}x_p}}{{\sqrt {{P_p}{{\left| {{g_{rp}}} \right|}^2} + {P_s}\bar h_r^H{{\bar h}_r} + \sigma _n^2} }}\\ &= \sqrt {{P_p}} {{\bar \omega }^H}{\hat H_j}{{\bar g}_p}x_p,
\end{split}
\end{equation}
\begin{equation}{}\nonumber
\begin{split}
{y^{(2)}_{j_{(s)}}} &= \mathop \sum \limits_{r=1}^R {\omega _r}{\hat h_{jr}}\frac{{\sqrt {{P_s}} {{\bar h}_r}\bar x_s}}{{\sqrt {{P_p}{{\left| {{g_{rp}}} \right|}^2} + {P_s}\bar h_r^H{{\bar h}_r} + \sigma _n^2} }}\\ &= \sqrt {{P_s}} {{\bar \omega }^H}{\hat H_j}{H_M} \bar x_s,
\end{split}
\end{equation}
\begin{equation}{}\nonumber
\begin{split}
{y^{(2)}_{j_{(n)}}} &= \mathop \sum \limits_{r=1}^R {\omega _r}{\hat h_{jr}}\frac{{{n_r}}}{{\sqrt {{P_p}{{\left| {{g_{rp}}} \right|}^2} + {P_s}\bar h_r^H{{\bar h}_r} + \sigma _n^2} }}+{n_j}\\ &= {{\bar \omega }^H}{\hat H_j}\bar n + {n_j},
\end{split}
\end{equation}
\end{subequations}
where
\begin{equation}{}\nonumber
\begin{split}
&\bar \omega  = {[\omega _1,\omega _2,  \ldots,\omega _R]^H}\nonumber,\\
&{\hat H_j} = diag({\hat h_{j1}},{\hat h_{j2}},  \ldots,  {\hat h_{jR}})\nonumber,\\
&{H_M} = [\overline h_1^T,\overline h_2^T,...,\overline h_R^T]^T\nonumber,\\
&{{\bar g}_p} = [{\tilde g}_ {1p},{\tilde g}_ {2p},\ldots,{\tilde g}_ {Rp}]^T\nonumber,\\
&\bar n = {[{\tilde n}_1,{\tilde n}_2 \ldots,{\tilde n}_R]^T}\nonumber
\end{split}
\end{equation}
and
\begin{equation}{}\nonumber
\begin{split}
&{\tilde g}_{rp} = \frac{{{g_{rp}}}}{{\sqrt {{P_p}{{\left| {{g_{rp}}} \right|}^2} + {P_s}\bar h_r^H{{\bar h}_r} + \sigma _n^2} }}\nonumber, \\
&{\tilde n}_r = \frac{{{n_r}}}{{\sqrt {{P_p}{{\left| {{h_{rp}}} \right|}^2} + {P_s}\bar h_r^H{{\bar h}_r} + \sigma _n^2} }}\nonumber
\end{split}
\end{equation}
for $r=1,2,...,R$. The coefficients' matrix of the channels between SU-TXs and relay nodes and the SU-TXs transmitted vector could be decomposed as follows
\begin{subequations}
\begin{equation}\nonumber{}
\begin{split}
&{H_{\rm{M}}} = [\underset{{H_{I1}:M_1}  \text{columns}}{\underbrace{{\underline h_1},...,{\underline h_{{{i}-1}}}}},{{\underline h}_{i}},{\underset{{H_{I2}:M_2}  \text{columns}}{\underbrace{\underline h_{{{i}+1}},...,{\underline h_M}}}}], \\
&\bar x = {[\underset{{x_{I1}:M_1}  \text{elements}}{\underbrace{x_1,...,x_{i-1}}},{x_{i}},\underset{{x_{I2}:M_2}   \text{elements}}{\underbrace{x_{i+1},..., x_M}}]^T},
\end{split}
\end{equation}
\end{subequations}
where $M_1 + M_2=M-1$. In (6a), $\underline h_i$, for $i=1,...,M$, indicates the channel vector between the $i$th SU-TX and the relay nodes.
Considering perfect channel state information (CSI) the signal power at the $j$th SU-RX could be written as,
\begin{equation}\nonumber{}
  S_j= {P_s} {{\bar \omega }^H}{\hat H_j}{{\underline h}_{j}}{{\underline h}_{j}^H}{\hat H_j}^H\bar \omega.
\end{equation}
Considering independent noise vectors on SU-RXs and relay nodes, the noise power at the $j$th SU-RX is
\begin{equation}{}
\begin{split}
N_j &= {{\bar \omega }^H}{\hat H_j}E\left\{ {\bar n{{\bar n}^H}} \right\}H_j^H\bar \omega  + \sigma _n^2 \nonumber
   \\ &= {{\bar \omega }^H}{\hat H_j}{{\rm{\Sigma }}_n}H_j^H\bar \omega  + \sigma _n^2. 
\end{split}
\end{equation}
Since the noise elements are considered to be independent on each receiver, $\Sigma_{n}$ is a diagonal matrix with elements of
\begin{equation}{}\nonumber
{\rm{\Sigma }}_n^{j,j} = \frac{{\sigma _n^2}}{{{P_p}{{\left| {{g_{rp}}} \right|}^2} + {P_s}\bar h_r^H{{\bar h}_r} + \sigma _n^2}}.
\end{equation}
Interference signal at each SU-RX consists of two parts. The former is the effect of PU-TX signal and the latter caused by undesirable SU-TXs' signal. They could be written as follows,\\
\begin{equation}{}
\begin{split}
       &{I_{j_{(p)}}^{(s)}} = {P_P}{{\bar \omega }^H}{\hat H_j}{{\bar g}_p}^T\bar g_p^*{\hat H_j}^H\bar \omega\nonumber,\\
    \\ &{I_{j_{(s)}}^{(s)}} ={P_s}{{\bar \omega }^H}{\hat H_j}{H_{I}}E\left\{ {  x_I}{x_I^H}\right\} {H_I^H{\hat H}}_j^H\bar \omega\nonumber
    \\ &\hspace{7mm}={P_s}{{\bar \omega }^H}{\hat H_j}{H_I}{H_I^H}{\hat H}_j^H\bar \omega, \nonumber
\end{split}
\end{equation}
where
\begin{equation}{}\nonumber
{{ {H}}_I} = \left[ {{{H}_{I1}},{ H_{I2}}} \right]\nonumber,\\
{x}_I = \left[ {x_{I1}},{x_{I2}} \right].\nonumber
\end{equation}
Accordingly, the SINR at the $j$th SU-RX is written as
\begin{equation}{}\nonumber
{SINR_j}  = \frac{{{P_s}{{\bar \omega }^H}{Q_{S_{j}}}\bar \omega }}{{{{\bar \omega }^H}\left( {{Q_{IP_{j}n}} + {Q_{IS_{j}n}} + {Q_{N_{j}}}} \right)\bar \omega + {c_j}}}
\end{equation}
where
\begin{equation}{}\nonumber
\begin{split}
    & {Q_{S_{j}}} = {\hat H_j}{{\bar h}_{j}}\bar h_{j}^H{\hat {H}_j}^H\nonumber,
 \\ & {Q_{IP_{j}}} = {P_P}{\hat H_j}{{\bar g}_p^T}\bar g_p^*{\hat H_j}^H\nonumber,
 \\ & {Q_{IS_{j}}} = {P_s}{\hat H_j}{H_I}{H_I^H}{\hat H_j}^H\nonumber,
 \\ & {Q_{N_{j}}} = {\hat H_j}{{\rm{\Sigma }}_n}{\hat H_j}^H\nonumber,
 \\ & c_j =\sigma _n^2.\nonumber
\end{split}
\end{equation}
The interference power at PU-RX could be written as,
\begin{equation}{}\nonumber
I_{(p)}^{(s)}= E{\left| {\mathop \sum \limits_{r = 1}^R {\hat g_{pr}}{\omega_r}{\frac{ { \sqrt {{P_s}} {{\bar h}_r}\bar x_s + {n_r}}}{{\sqrt {{P_p}{{\left| {{g_{rp}}} \right|}^2} + {P_s}\bar h_r^H{{\bar h}_r} + \sigma _n^2} }}}}\right|^2}.
\end{equation}
Hence
\begin{equation*}{}
I_{(p)}^{(s)}={\rm{  }}{{\bar \omega }^H}{\hat H_p}({P_s}{H_M}H_M^H + {{\rm{\Sigma }}_n}){\hat H_p}^H\bar \omega,
\end{equation*}
where
\begin{equation}{}\nonumber
{\hat H_p} = diag({\tilde g}_{p1},{\tilde g}_{p2}, \ldots, {\tilde g}_{pR}),
\end{equation}
and
\begin{equation}{}\nonumber
{\tilde g}_{pr} = \frac{{{\hat g_{pr}}}}{{\sqrt {{P_p}{{\left| {{g_{rp}}} \right|}^2} + {P_s}\bar h_r^H{{\bar h}_r} + \sigma _n^2}, }},\hspace {5mm} r=1,...,R.
\end{equation}
\section{Beamforming Design}
According to the derived expressions for SINR at SU-RX and interference at PU-RX, the optimization problem with the objective of maximizing the minimum SINR at the SU-RXs with the constraints of interference on PU-RX and the relay network transmitted power could be written as follows,
\begin{equation}{}\label{eq033}
\begin{split}
   \underset{\bar\omega}{\text{max}}\hspace{2mm}\ \text{min} \ &  \qquad  \{SINR_1,SINR_2,...,SINR_N\} \nonumber\\
   \text{s.t}\ &\\
   \ &  I_{(p)}^{(s)} < {I_p}\nonumber,\\
   \ & \mathop \sum \limits_{r=1}^R {\left| {{\omega _r}} \right|^2} < {P_t},\nonumber
\end{split}
\end{equation}
where $I_p$ and $P_t$ are the interference limit on the PU-RX and the relay networks budget for transmitted power , respectively. Now, by introducing the new parameter $\rho$, which plays the role of minimum received SINR at the SU-RXs, the optimization problem could be written as
\begin{equation}{}\label{eq033}
\begin{split}
   \underset{\bar\omega}{\text{max}} \ &  \qquad \qquad \rho \nonumber \\
   \text{s.t}\\
   \ &   \frac{{{P_s}{{\bar \omega }^H}{Q_{S_{j}}}\bar \omega }}{{{{\bar \omega }^H}\left( {{Q_{IP_{j}}} + {Q_{IS_{j}}} + {Q_{N_{j}}}} \right)\bar \omega  + {c_j}}} \ge {\rho ^2}, \hspace{5mm} j=1,...,N. \nonumber \\
    \ & I_{(p)}^{(s)} < {I_p}\nonumber \\
   \ & \mathop \sum \limits_{r=1}^R {\left| {{\omega _r}} \right|^2} < {P_t}.\nonumber
\end{split}
   \end{equation}
Then it could be reformulated as SOCP problem
\begin{equation}\tag{4}
\begin{split}
  \underset{\bar\omega}{\text{max}} &  \qquad \qquad \rho \\
   \text{s.t}\\
   \ & \bar \omega _1^H{ Q_{S1_{j}}} \ge \frac{\rho }{{\sqrt {{P_s}} }}\parallel \bar \omega _1^H Q_I^{\frac{1}{2}}{\parallel _2}\\
   \ & \sqrt {{I_p}}  \ge \parallel \bar \omega _1^H{\hat H_{q1}}^H{Q}_r^{1/2}{\parallel _2} \\
   \ & {\rm{ }}tr(\bar \omega ^H{{\bar \omega }})
\end{split}
\end{equation}
where
\begin{equation}{}\nonumber
\begin{split}
    & {{\bar \omega }_1} = {\left[ {1,{{\bar \omega }^T}} \right]^T}\nonumber,
 \\ & {Q_{S1_{j}}} = {[0,\{(Q_{S_{j}})^\frac{1}{2}\}^T]^T}\nonumber,
 \\ & {Q_I} = \left[ {\begin{array}{*{20}{c}}
{{c_j}}&{{0^{1 \times R}}}\\
{{0^{R \times 1}}}&{{Q_{IP_{j}}} + {Q_{IS_{j}}} + {Q_{N_{j}}}}\nonumber
\end{array}} \right],
 \\ & {{\hat H}_{q1}} = \left[ {\begin{array}{*{20}{c}}
0&{{0^{1 \times R}}}\\
{{0^{R \times 1}}}&{{\hat H_q}}\nonumber
\end{array}} \right],
 \\ & {{Q}_r} = \left[ {\begin{array}{*{20}{c}}
1&{{0^{1 \times R}}}\\
{{0^{R \times 1}}}&{{Q_s} + {{\rm{\Sigma }}_n}}\nonumber
\end{array}} \right],
\end{split}
\end{equation}
 and
 \begin{equation}{}\nonumber
{{Q_s} = {P_s}{H_M}H_M^H}.
\end{equation}
For any fixed value of $\rho$ the SOCP problem could be written as
 \begin{equation}{}\nonumber
 \begin{split}
  \text{find} \ &  \qquad \qquad \bar\omega \\
   \text{s.t}\\
   \ & {\bar \omega}^H Q_{S1_{j}} \ge \frac{\rho}{\sqrt{P_s}} \parallel{{\bar \omega_1}}^H \tilde {Q}_I^{\frac{1}{2}}\parallel_2\\
   \ &  \sqrt{I_P} \ge \parallel{{\bar \omega _{1}}}^H  {\hat H}^H_{q1} Q^{\frac{1}{2}}_r\parallel_2\\
   \ &  tr({\bar \omega}^H {\bar \omega}^H ) \le P_t.
\end{split}
   \end{equation}
This problem is a SOCP feasibility problem and its feasibility depends on the optimal solution of (4). Therefore, considering $\rho_{0}$ is the optimal solution of (4), the SOCP feasibility problem has a solution only if ${\rho}<{\rho_{0}} $. In order to find this optimal solution a bi-section search method could be used \cite{boyd2004convex}.

\section{Simulation Results}
In our simulations, the transmitted power of both PU-TX and the SU-TX are assumed to be $5$ dB. In Fig. 2, assuming $R=10$ and $M=N=3$, the worst case SINR variations versus total relay transmitted power is showed for 3 different values of interference limit on PU-RX. As it is shown, the received SINR at the SU-RX improves as the total relay transmitted power increases. On the other hand, by increasing the interference limit on PU-RX from $-5$ dB to $0$ dB, performance improvement of the system is slightly less than the improvement when it increases form $-10$ dB to $-5$ dB. However, by increasing the number of secondary pairs, interference power on each SU-RX is also increased. Thus, the worst case SINR decreases if the number of secondary pairs increase, as it is also shown in Fig. 3, when $R=10$ and $I_p=10$ dB. Finally, when $M=N=3$ and $I_p=0$ dB, Fig. 4 illustrates that the system performance improves by the number of relay nodes. However, this improvement is diminishes as the number of relays increases. The diminishing return of the relay nodes is also in line with intuition, since as they increase the interference power on SU-RXs also increase, which is a reduces the performance improvement.
\begin{figure}[!h]
   \begin{center}
    \includegraphics[width=\columnwidth]{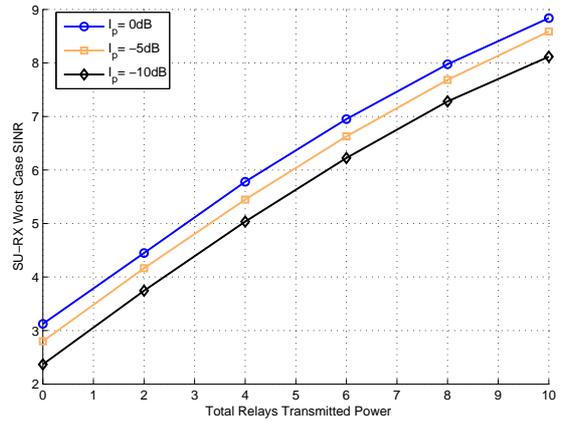}
    \caption{The SU-RXs worst case SINR versus total relays transmitted power, for different values of interference limit on PU-RX.}\label{figsystemmodel}
     \end{center}
\end{figure}

\begin{figure}[!h]
   \begin{center}
    \includegraphics[width=\columnwidth]{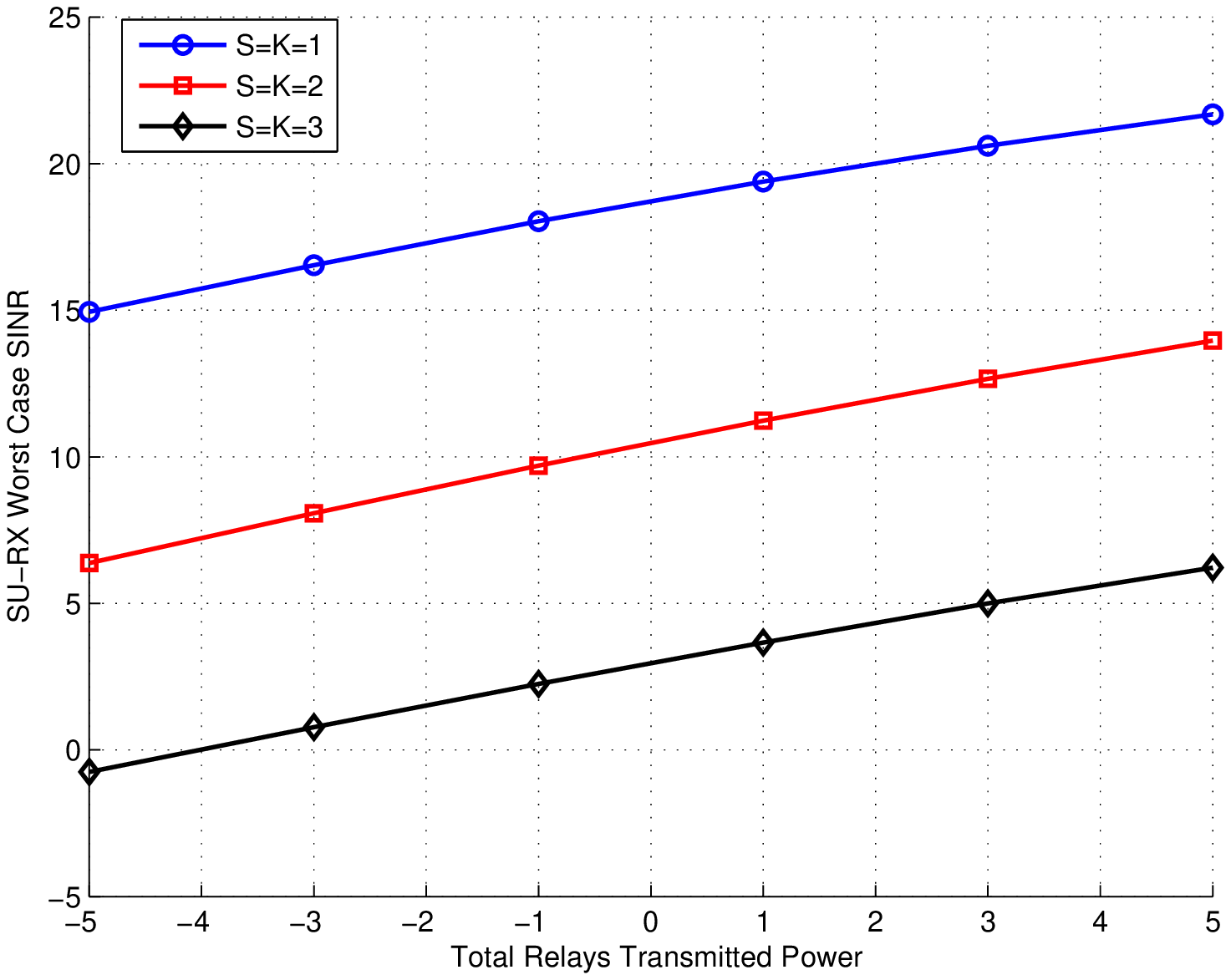}
    \caption{The SU-RXs worst case SINR versus total relays transmitted power, for different number of secondary pairs.}\label{figsystemmodel}
     \end{center}
\end{figure}
   
\begin{figure}[!h]
   \begin{center}
    \includegraphics[width=\columnwidth]{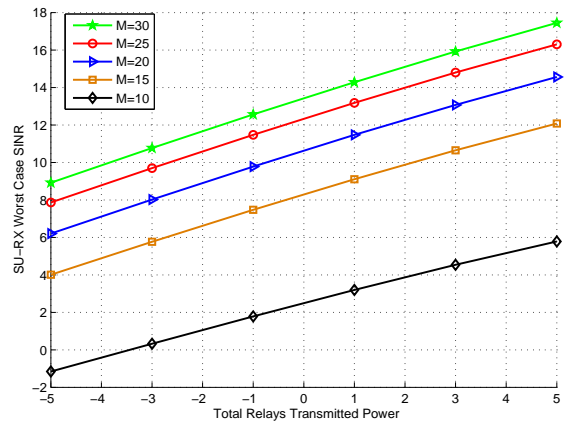}
    \caption{The SU-RXs worst case SINR versus total relays transmitted power, for different number of relays.}\label{figsystemmodel}
     \end{center}
\end{figure}

\section{Conclusion}
A beamforming problem in a cognitive radio relay network was studied. The beamforming coefficients were drawn from the solution of a convex optimization problem, with maximizing the worst case SINR among SU-RXx as objective and total relay network transmit power and limited interference at PU-RX as constraints. The optimization problem was reformulated as a SOCP problem and was solved by a bi-section search method. The result of our simulation shows the behavior of performance improvement as a function of the size and the power of the relay network.

\bibliographystyle{ieeetr}
\bibliography{ref}

\begin{thebibliography}{10}

\bibitem{goldsmith2009breaking}
A.~J. Goldsmith, S.~A. Jafar, I.~Maric, and S.~Srinivasa, ``Breaking spectrum
  gridlock with cognitive radios: An information theoretic perspective.,'' {\em
  Proceedings of the IEEE}, vol.~97, no.~5, pp.~894--914, 2009.

\bibitem{zheng2010robust}
G.~Zheng, S.~Ma, K.-K. Wong, and T.-S. Ng, ``Robust beamforming in cognitive
  radio,'' {\em IEEE Transactions on Wireless communications}, vol.~9, no.~2,
  pp.~570--576, 2010.

\bibitem{piltan2012distributed}
A.~Piltan and S.~Salari, ``Distributed beamforming in cognitive relay networks
  with partial channel state information,'' {\em IET communications}, vol.~6,
  no.~9, pp.~1011--1018, 2012.

\bibitem{du2011joint}
H.~Du, T.~Ratnarajah, M.~Pesavento, and C.~B. Papadias, ``Joint transceiver
  beamforming in mimo cognitive radio network via second-order cone
  programming,'' {\em IEEE Transactions on Signal Processing}, vol.~60, no.~2,
  pp.~781--792, 2011.

\bibitem{chen2009filter}
H.~Chen, A.~B. Gershman, and S.~Shahbazpanahi, ``Filter-and-forward distributed
  beamforming in relay networks with frequency selective fading,'' {\em IEEE
  Transactions on Signal Processing}, vol.~58, no.~3, pp.~1251--1262, 2009.

\bibitem{mohammadkhani2014robust}
S.~Mohammadkhani, S.~M. Razavizadeh, and I.~Lee, ``Robust filter and forward
  relay beamforming with spherical channel state information uncertainties,''
  in {\em 2014 IEEE International Conference on Communications (ICC)},
  pp.~5023--5028, IEEE, 2014.

\bibitem{mohammadkhani2012cooperative}
S.~Mohammadkhani, M.~H. Kahaei, and S.~M. Razavizadeh, ``Cooperative
  filter-and-forward beamforming in cognitive radio relay networks,'' in {\em
  6th International Symposium on Telecommunications (IST)}, pp.~170--175, IEEE,
  2012.

\bibitem{beigi2009cooperative}
M.~A. Beigi and S.~M. Razavizadeh, ``Cooperative beamforming in cognitive radio
  networks,'' in {\em 2009 2nd IFIP Wireless Days (WD)}, pp.~1--5, IEEE, 2009.

\bibitem{bayat2011cognitive}
S.~Bayat, R.~H. Louie, Y.~Li, and B.~Vucetic, ``Cognitive radio relay networks
  with multiple primary and secondary users: Distributed stable matching
  algorithms for spectrum access,'' in {\em 2011 IEEE International Conference
  on Communications (ICC)}, pp.~1--6, IEEE, 2011.

\bibitem{boyd2004convex}
S.~Boyd and L.~Vandenberghe, {\em Convex optimization}.
\newblock Cambridge university press, 2004.

\end{thebibliography}

\end{document}